\documentclass[aps,prd,twocolumn,showpacs,superscriptaddress,preprintnumbers,floatfix,nofootinbib,10pt]{revtex4-2}
\usepackage{graphicx}
\usepackage{amsmath, amsthm, amssymb}
\usepackage{slashed}
\usepackage{url}
\usepackage[dvipsnames]{xcolor}
\usepackage{bm}
\usepackage{mathtools}
\usepackage[normalem]{ulem}
\usepackage{comment}
\usepackage{siunitx}
\usepackage{multirow}
\usepackage{natbib}
\usepackage[colorlinks=true,citecolor=blue,urlcolor=blue,linktocpage=true,linkcolor=blue]{hyperref}
\usepackage[capitalize]{cleveref}

\newcommand{\Th}{$^{229}$Th }

\newcommand{\Mpl}{M_{\text{P}}}

\newcommand{\bea}{\begin{eqnarray}}
\newcommand{\eea}{\end{eqnarray}}

\newcommand{\dnuDM}{\delta \nu_{\mathrm{DM}}}
\newcommand{\mDM}{m_{\mathrm{DM}}}
\newcommand{\TDM}{T_{\mathrm{DM}}}

\begin{document}
\preprint{KEK-QUP-2026-0003, DESY-26-022}

\title{
Probing Ultralight Dark Matter at the Mega-Planck Scale with the Thorium Nuclear Clock
}

\author{Jason Arakawa}
\email{arakawaj@udel.edu}
\affiliation{Department of Physics and Astronomy, University of Delaware, Newark, Delaware 19716, USA}
\affiliation{International Center for Quantum-field Measurement Systems for Studies of the Universe and Particles (QUP, WPI),
High Energy Accelerator Research Organization (KEK), Oho 1-1, Tsukuba, Ibaraki 305-0801, Japan}

\author{Jack F. Doyle}
\affiliation{JILA, NIST and University of Colorado, Department of Physics, University of Colorado, Boulder, CO 80309}

\author{Elina~Fuchs}
\email{elina.fuchs@desy.de}
\affiliation{Deutsches Elektronen-Synchrotron DESY, Notkestr. 85, 22607 Hamburg, Germany}
\affiliation{Institut f\"{u}r Theoretische Physik, Leibniz Universit\"{a}t Hannover, Appelstraße 2, Hannover, 30167, Germany}

\author{Jacob S. Higgins}
\affiliation{JILA, NIST and University of Colorado, Department of Physics, University of Colorado, Boulder, CO 80309}

\author{Fiona~Kirk}
\email{fiona.kirk@weizmann.ac.il}
\affiliation{Department of Particle Physics and Astrophysics, Weizmann Institute of Science, Rehovot 761001, Israel}
\affiliation{Physikalisch-Technische Bundesanstalt, Bundesallee 100, Braunschweig, 38116, Germany}
\affiliation{Institut f\"{u}r Theoretische Physik, Leibniz Universit\"{a}t Hannover, Appelstraße 2, Hannover, 30167, Germany}

\author{Kai Li}
\affiliation{JILA, NIST and University of Colorado, Department of Physics, University of Colorado, Boulder, CO 80309}

\author{Tian Ooi}
\affiliation{JILA, NIST and University of Colorado, Department of Physics, University of Colorado, Boulder, CO 80309}

\author{Gilad~Perez}
\email{gilad.perez@weizmann.ac.il}
\affiliation{Department of Particle Physics and Astrophysics, Weizmann Institute of Science, Rehovot 761001, Israel}

\author{Wolfram~Ratzinger} 
\email{wolfram.ratzinger@weizmann.ac.il}
\affiliation{Department of Particle Physics and Astrophysics, Weizmann Institute of Science, Rehovot 761001, Israel}

\author{Marianna S. Safronova}
\affiliation{Department of Physics and Astronomy, University of Delaware, Newark, Delaware 19716, USA}

\author{Thorsten~Schumm}
\email{thorsten.schumm@tuwien.ac.at}
\affiliation{Vienna Center for Quantum Science and Technology, Atominstitut, TU Wien, Stadionallee~2, 1020 Wien, Austria}

\author{Jun Ye}
\email{ye@jila.colorado.edu}
\affiliation{JILA, NIST and University of Colorado, Department of Physics, University of Colorado, Boulder, CO 80309}

\author{Chuankun Zhang}
\affiliation{JILA, NIST and University of Colorado, Department of Physics, University of Colorado, Boulder, CO 80309}

 
\date{\today}
\begin{abstract}
Ultralight dark matter is expected to induce oscillations of nuclear parameters. 
These oscillations are characterized by extremely weak couplings or high suppression scales, with the Planck scale -- the characteristic scale of quantum gravity -- serving as a natural benchmark. 
Probing this phenomenon requires systems with exceptional sensitivity to shifts in nuclear energies. 
The uniquely low-energy nuclear isomeric transition in \Th provides such sensitivity: it directly probes the nuclear interaction and, owing to a near cancellation between electromagnetic and nuclear contributions,  its response to changes in nuclear structure is greatly amplified.
We devise and perform a new type of ultrasensitive search for dark matter which uses the precision nuclear spectroscopy at JILA to set the strongest bounds in the mass range $10^{-21}\,{\rm eV} \lesssim \mDM \lesssim 10^{-19}\,{\rm eV}$. Our results probe effective interaction scales exceeding $10^6$ times the Planck scale (the Mega-Planck scale) and establish the \Th system as the leading probe of dark matter couplings to the nuclear sector.
\end{abstract}

\maketitle


\section{Introduction}
\begin{figure*}
    \centering
    \includegraphics[width=0.99\linewidth]{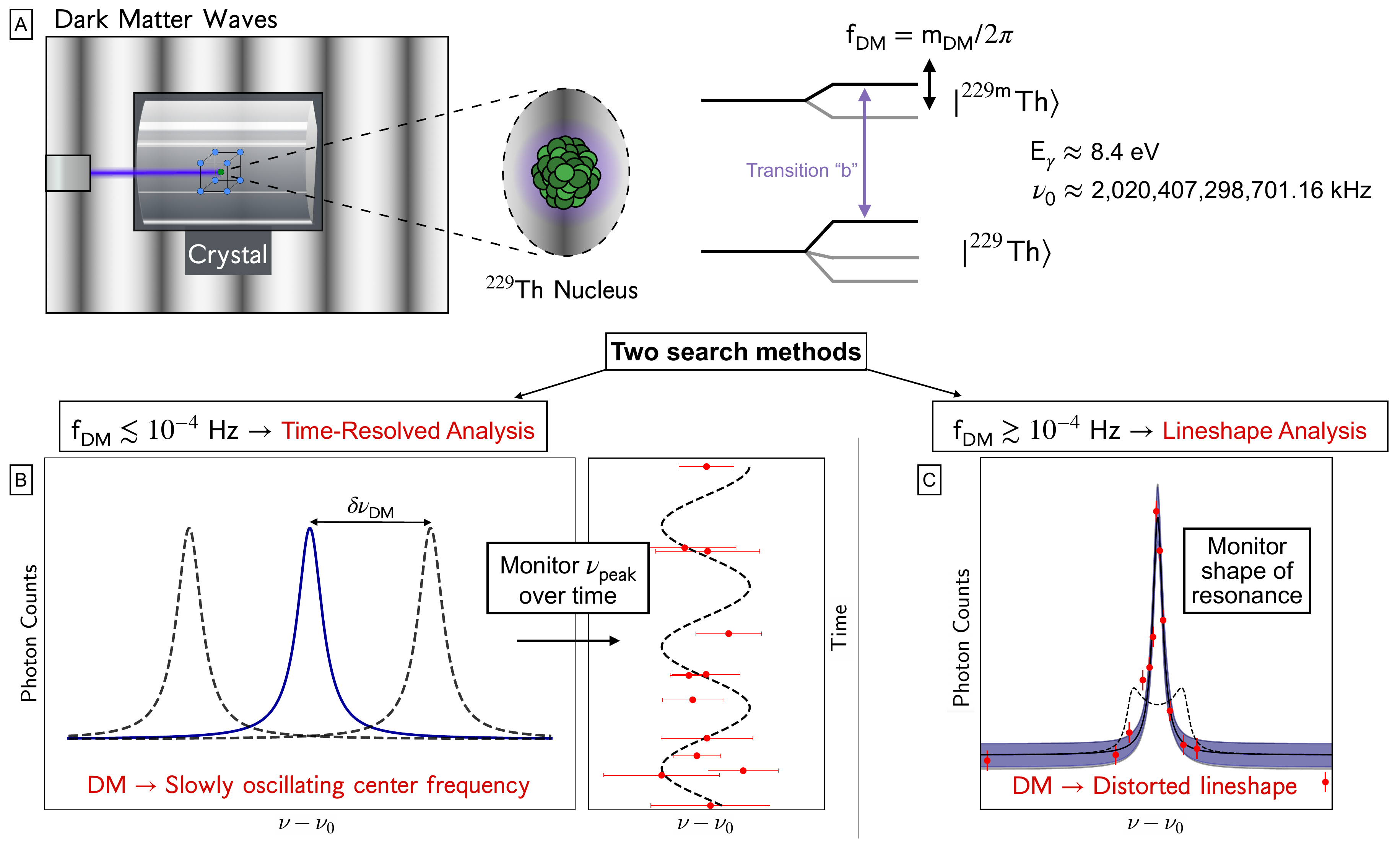}
    \caption{Schematic diagram of the influence of dark matter (DM) on the $^{229}$Th spectroscopy, leading to two search methods relevant for different DM frequency regimes. (A) A cartoon of thorium nuclei embedded  in a crystal, in the presence of ULDM waves (dark gray bands). The isomer transition is also shown. (B) A schematic of the time-resolved analysis, which relies on the varying position of the centre frequency of the transition, as relevant for low-mass (frequency) DM. (C) The lineshape analysis, where the DM oscillation frequencies are fast enough such that the shape of the resonance is changed, either through broadening or in the most extreme circumstances, a double peak structure appears.
    }
    \label{fig:pedagogical_intro}
\end{figure*}

Dark matter (DM) constitutes most of the matter in the Universe, yet its microscopic nature remains unknown. 
Ultralight dark matter (ULDM) is arguably the simplest class of DM candidates and is naturally produced via the misalignment mechanism in the early Universe~\cite{Preskill:1982cy,Abbott:1982af,Dine:1982ah}.
Many well-motivated ULDM models predict that the dominant non-gravitational interactions of DM are not with electromagnetism, but with the strong nuclear sector of the Standard Model. More specifically, these interactions are with neutrons and protons (or, more fundamentally, with their quark and gluon constituents).
Representative examples for ULDM models with couplings to the strong nuclear sector include the QCD axion~\cite{Preskill:1982cy,Abbott:1982af,Dine:1982ah}, the dilaton~\cite{Arvanitaki:2014faa} (see, however,~\cite{Hubisz:2024hyz}), relaxion models~\cite{Graham:2015ifn,Banerjee:2018xmn}, Higgs portals~\cite{Piazza:2010ye}, and alternatives to the QCD axion~\cite{Dine:2024bxv}. 
In these scenarios, the interactions between DM and the Standard Model are suppressed by very large energy scales, rendering them exceptionally weak and challenging to probe experimentally. 

More generally, regardless of any specific DM theory, the Planck scale provides an important target, since it is the characteristic scale at which gravity becomes strong and new fundamental physics is expected to emerge.
In this work, we break new ground by probing couplings between ULDM and the electromagnetic and nuclear fields that are suppressed by an energy six orders of magnitude above this scale.

ULDM behaves as a classical background field, oscillating at a frequency set by its mass $\mDM$.
Such oscillations induce time variations of the fundamental constants via the ULDM couplings (see~\cite{Antypas:2022asj} for a recent review and references therein).
In particular, such couplings lead to a modulation of transition frequencies 
\begin{align}
    \nu(t) \simeq \nu_0 + \dnuDM \cos\left(\mDM t\right)\,,
    \label{eq:nuosc}
\end{align}
where the amplitude $\dnuDM$ depends on the local DM density, the strength of the DM couplings and the sensitivity of the transition in question. 

In this work, we use recent and new high-resolution spectroscopic data of the \Th isomer transition acquired at JILA, which is recorded via direct frequency comparison with a Sr-based optical atomic clock~\cite{zhang_frequency_2024,ooi_frequency_2025}. The $^1S_0 – ^3P_0$ transition of the Sr clock has one of the lowest sensitivities to changes of fundamental constants and thereby serves as an anchor frequency reference~\cite{Ludlow:2008_Sr, Blatt:2008_Sr}.
In stark contrast, the uniquely low-energy ($\sim\SI{8}{eV}$) nuclear transition in \Th, which is accessible to laser spectroscopy~\cite{Tiedau:2024obk}, exhibits an extraordinary $10^{8}-10^{10}$ enhancement in sensitivity to variations of nuclear parameters compared to optical transitions~\cite{Fuchs:2024xvc,Beeks:2024xnc}.
This combination of an exceptionally insensitive reference and an exceptionally sensitive probe enables us to surpass existing searches based on optical atomic clocks and tests of violation of the equivalence principle, even before the nuclear clock has reached atomic clock-level precision.

A key novelty of this work is a unified analysis strategy applied to time-stamped nuclear spectroscopy data recorded over ten months, with individual scans lasting approximately two hours (see Fig.~\ref{fig:pedagogical_intro}). 
A \emph{time-resolved analysis} of the transition frequency searches for slow DM-induced oscillations of the transition frequency, with periods ranging from months down to hours. A refined \emph{lineshape analysis}, building on Ref.~\cite{Fuchs:2024xvc}, probes faster oscillations that occur within a single line scan by detecting DM-induced distortions of the spectral profile. Applied to new and recent spectroscopic data, this combined strategy improves sensitivity by six to seven orders of magnitude compared to Ref.~\cite{Fuchs:2024xvc}.

\section{Data Taking}
\label{sec:exp}
High-resolution frequency-based precision laser spectroscopy~\cite{zhang_frequency_2024,ooi_frequency_2025} was performed in JILA on three $^{229}$Th-doped CaF$_2$ crystals (C10, C13, X2) produced by TU Wien~\cite{beeks_growth_2023}, each with different doping concentrations. The nuclear transition in these crystals is split via the interaction between the nuclear quadrupole moment and the electric field gradient the nuclei experience in the crystal environment, resulting in five main spectral lines. 
Following the designation given in Ref.~\cite{zhang_frequency_2024}, the line ``b" $(m_g=\pm\frac{5}{2} \rightarrow m_e=\pm\frac{3}{2})$ exhibits the narrowest linewidth and the smallest coefficient of the temperature dependence of its center frequency~\cite{ooi_frequency_2025}, and is thus used for the model constraint reported in this paper.

The data were taken with a vacuum ultraviolet (VUV) frequency comb~\cite{zhang_tunable_2022} referenced to the JILA Sr optical atomic clock~\cite{aeppli_clock_2024} ensuring precise and accurate frequency reproducibility. By precisely tuning either the comb carrier-envelope offset frequency or the repetition frequency, we can scan the narrowline comb teeth over any spectral feature of interest within the spectral envelope of the comb. 

When a comb tooth is scanned over the nuclear transition, it is held at a fixed frequency to excite nuclei for $t_e =$ 400~seconds; nuclear fluorescence is subsequently spectrally filtered and collected with a photomultiplier tube for $t_d =$ 200\,s, then a new frequency step is taken. About $N_{\rm pts}\approx 13$ are taken for a specific resonance, and thus the lineshape of each spectral component is typically taken over $T_{\rm scan} \approx N_{\rm pts} (t_e + t_d)\approx$ 2~hours. The data is post-processed to account for systematic effects such as residual fluorescence from previous excitations and fluctuations in laser intensity. A Lorentzian lineshape is found to better describe our spectroscopy data than a Gaussian~\cite{ooi_frequency_2025,girvin_prospects_2025}; therefore, it is used to extract the line center.

The center frequency of line “b” has a vanishing first order dependence on temperature at 195\,K. With temperature stabilization at 195(1)\,K via a servo controlled heater and a liquid nitrogen bath~\cite{higgins_temperature_2025}, the line b center frequency recorded over a year is consistent at the kHz level across the three crystals. For the two lower doping concentration crystals (C10 and C13), their line centers are mutually consistent and reproducible at 200 Hz over a course of $T_{\rm tot} \approx $ 10\,months~\cite{ooi_frequency_2025}.  

The comb tooth linewidth is estimated to be below 1\,kHz.  The spectroscopic linewidths are dominated by inhomogeneous broadening within the crystal due to lattice-mediated interactions between dopant sites with inhomogeneous inter-site distance~\cite{girvin_prospects_2025}. The linewidth of each spectral component grows linearly with the Th doping concentration, with line “b” exhibiting the smallest linear slope and showing an extrapolated linewidth of about 20 kHz at zero concentration.  
No measurable effect of temperature on the transition linewidth is observed~\cite{ooi_frequency_2025}. 

\section{Data Analysis}
\label{sec:data_analysis}
Using the time-stamped laser-spectroscopy data acquired in Ref.~\cite{ooi_frequency_2025} and beyond, we employ two complementary methods to constrain variations of fundamental constants, depending on the desired search range of the DM mass and the corresponding oscillation frequency.
In Section~\ref{sec:timeresolved}, we perform a time-resolved analysis of the \Th  ``b" line transition frequency, treating the data analogously to atomic clock-comparison tests and searching for periodic signals by means of the Lomb-Scargle periodogram. 
Since acquiring a single spectrum takes about $T_{\rm scan}\sim 2$\,hours, this approach is sensitive only to slow frequency variations on timescales ranging from months down to hours.
To constrain faster oscillations, we perform a lineshape analysis, as described in Section~\ref{sec:lineshape-analysis}. Rapid frequency variations distort or broaden the spectral line, enabling sensitivity to oscillations beyond the time resolution of individual scans. Building on Ref.~\cite{Fuchs:2024xvc}, we develop both a conservative, profile-agnostic approach and a more refined, physics-motivated lineshape analysis, and explicitly assess the systematic uncertainties by comparing the different approaches.  

\subsection{Time-resolved analysis}
\label{sec:timeresolved}

We search for sinusoidal oscillations in the centre frequency $\nu(t)$ of the $^{229}{\rm Th}$ isomer transition, closely following the technique used in atomic-clock comparisons~\cite{VanTilburg:2015oza, Hees:2016gop, Sherrill:2023zah, Filzinger:2023zrs}. The frequency of DM-induced oscillations is set by the DM mass, $\omega = 2\pi f_{\rm DM} = \mDM$. Since the phase of the DM oscillation is unknown, the time-dependent transition frequency from \cref{eq:nuosc} can be expressed as $\nu(t) = \nu_{0} + a \cos{(\mDM t)} + b \sin{(\mDM t)}$, where $\nu_{0}$ is the central frequency in the absence of DM. The total modulation amplitude is $\dnuDM = \sqrt{a^2 + b^2}$. 

Within this method, the range of DM masses $m_{{\rm DM}}$ that can be probed by the data set is $2\pi/T_{\rm tot} \lesssim  \mDM \lesssim \pi/T_{\rm scan}$, 
where $T_{\rm tot}\approx 10$\,months is the total time the data was recorded, and $T_{\rm scan} \approx 2$ hours is the time it takes to record one spectrum.

We search for periodic modulations of the transition frequency using a Lomb–Scargle periodogram applied to the time-stamped spectroscopy data, supplemented by additional scans (see Supplementary Information (SI) for an overview). Upper limits on the modulation amplitude are derived from Monte Carlo simulations incorporating the measured frequency uncertainties. To account for the look-elsewhere effect across the scanned mass range, we determine a global false-alarm threshold from the distribution of maximal amplitudes obtained in the simulations (see SI for details). No statistically significant periodic signal is observed in the range $10^{-21}\,{\rm eV}\leq \mDM\leq 3\times10^{-19}\,{\rm eV}$\,.

As a cross-check, we perform direct sinusoidal fits at fixed dark-matter mass and obtain consistent bounds. We further verify the robustness of the results against dataset selection. 
In the main analysis, we adopt the most stable dataset (line “b” from the two low-doping crystals at 195\,K), which minimizes systematic uncertainties. Including additional data strengthens the bounds but introduces larger systematics (see SI).

\subsection{Lineshape analysis}
\label{sec:lineshape-analysis}
If the DM background oscillates multiple times during a single scan of duration $T_{\rm scan}$, the time-resolved approach is no longer applicable. Instead, rapid frequency modulation alters the resonance profile itself~\cite{Fuchs:2024xvc}. 
When the modulation amplitude exceeds the intrinsic linewidth, the line broadens or even splits, providing a distinct experimental signature.

In the absence of intrinsic broadening, periodic modulation, averaged over at least one DM cycle, would produce an arcsine distribution supported on the interval $\left[ \nu_0-\dnuDM, \nu_0 +\dnuDM\right]$~\cite{Fuchs:2024xvc}. 
In practice, the nuclear resonance exhibits inhomogeneous broadening due to crystal strain, the finite linewidth of the laser beam used to excite the thorium nuclei, thermal fluctuations, and the natural linewidth itself.
Although this broadening cannot be predicted from first principles, the qualitative intuition regarding DM-induced lineshape distortions remains valid.

We therefore implement two complementary search strategies. First, a conservative, width-based approach interprets the observed linewidth as entirely arising from DM-induced molulation, yielding the robust bound
\begin{align}
    \delta \nu_{\rm DM} \lesssim \frac{1}{2}\Gamma_{\rm agn}=8.2\,\mathrm{kHz}\,,
    \label{eq:conservative_UB}
\end{align}
where $\Gamma_{\rm agn}$ denotes the linewidth, estimated in a model-independent manner (see the SI for details). 
The corresponding half-width-based bound is shown in Figures~\ref{fig:scalar_nuclear_bounds},\ref{fig:de_bound}, and \ref{fig:axion_bound} and labelled ``width''.

Second, the physics-informed lineshape analysis simultaneously fits the background profile and the DM-induced frequency modulation, while taking into account the time dependence of the experimental procedure (see Section~\ref{sec:exp}).
For the dilute thorium concentration in the CaF${}_2$ crystals used for this study, defect-induced long-range electric-field gradient inhomogeneities produce inhomogeneous broadening with power-law tails, resulting in a Lorentzian lineshape~\cite{CohenReif1957,stoneham_shapes_1969,girvin_prospects_2025,ooi_frequency_2025}. 
We therefore model the intrinsic lineshape as Lorentzian, which closely matches the measured profile and allows DM-induced deviations to be more tightly constrained than when applying the conservative width-based method.

The difference between the bounds obtained using the first and the second method is used to estimate systematic uncertainties.
To cross-check the dependence of the lineshape-based bounds on the background lineshape profile, we compare them with the lineshape bounds obtained using Gaussian and Crystal Ball profiles (see the SI).

Goodness of fit and confidence intervals are estimated via Monte Carlo sampling of the model parameters, with predicted photon counts compared to the data using a log-likelihood function (see SI for details). 
This analysis is repeated for DM masses $\mDM$ ranging from $\mDM = 2\pi/T_{\rm scan} 
\approx 6\times 10^{-19}$\,eV -- where one scan probes approximately one DM oscillation -- to  $\mDM \approx 4\times 10^{-15}$\,eV
for which each data point averages over $\mathcal{O}(10^2)$ oscillations.

Both the bound obtained via the agnostic width as well as the one from the lineshape analysis at large masses can be extrapolated to dark matter masses $\mDM\gg 2\pi /t_e$\,. This holds true as long as the dark matter modulation frequency $f_{\rm DM}=\mDM/(2\pi)$ is smaller than the amplitude of the oscillations $f_{\rm DM}\lesssim\dnuDM$\,. For $\dnuDM\lesssim f_{\rm DM}$ the frequency modulation induced by the ULDM manifests in the appearance of sidebands, rather than a broadening of of the line, which would require a dedicated search \cite{Fuchs:2024xvc}. Our lineshape based bounds therefore apply up to masses comparable to the linewidth, i.e. $\mDM\sim 7\times 10^{-11}\,{\rm eV}\,(f_{\rm DM} \sim 16{\rm kHz})$\,.
\subsection{Translation into bounds on couplings} 
\label{sec:bounds}

\begin{figure*}[ht]
    \includegraphics{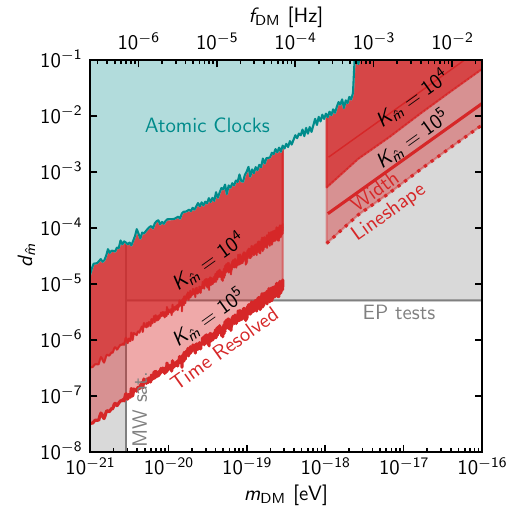}
    \includegraphics{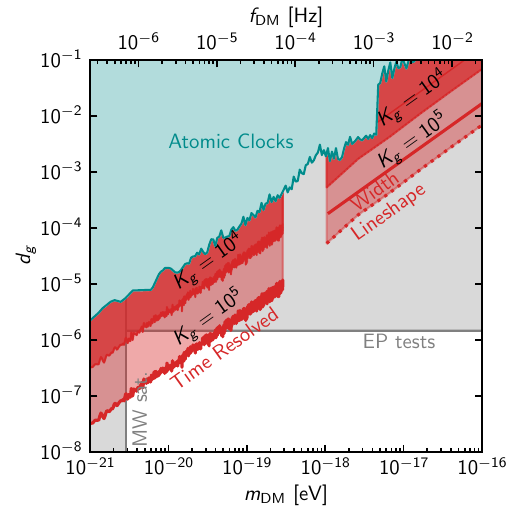}
    \caption{%
        Searches for coupling of scalar DM $\phi$ to quark masses (left) and gluons (right). The coloured regions represent constraints from searches for time variations in frequencies, assuming the scalar field constitutes DM. 
        The red region shows the constraint derived in this work under two scenarios for the sensitivity factor $K$ (see text for details on the challenge to calculate $K$). 
        For small masses and correspondingly low frequencies, the bound is inferred from the time-resolved analysis. For larger masses, we apply the lineshape analysis. For the lineshape analysis we further show an agnostic bound relying purely on the recorded width (straight) as well as one where a Lorentzian shape arising from the crystal environment is assumed (dotted). 
        In contrast, the gray regions are excluded by constraints that rely on the scalar being sufficiently massive to support the existence of Milky Way (MW) satellite galaxies~\cite{DES:2020fxi}, as well as from tests of equivalence principle (EP) violations mediated by the scalar field~\cite{MICROSCOPE:2022doy}.
       The provisional operation of a nuclear clock already bests the bounds coming from atomic clocks reaching the standard quantum limit~\cite{Hees:2016gop,Kennedy:2020bac,Kobayashi:2022vsf,Sherrill:2023zah,Filzinger:2023zrs,Banerjee:2023bjc} (teal). At masses $\mDM\lesssim 10^{-19}$ the nuclear clock already explores yet unconstrained territory. 
        }
    \label{fig:scalar_nuclear_bounds}
\end{figure*}

As mentioned above, nuclear clocks are much more sensitive than atomic clocks to effects that modify nuclear physics. We therefore restrict the discussion here to a dilatonic scalar that exclusively couples to the nuclear sector, and comment on its couplings to electromagnetism in the SI. 
Above the quantum chromodynamics (QCD) confinement scale, $\Lambda_{\rm QCD}$, the couplings of such a scalar field $\phi$ are described by~\cite{Arvanitaki:2014faa}

\begin{align} 
\label{eq:scalar_Lint}
    \mathcal{L}_{\text{int}} =  
    -\kappa \phi
    \bigg( \frac{d_g \beta(g)}{2g}G_{\mu\nu}^A G^{A\mu\nu}+\sum_{q}(d_{m_q}+\gamma_{m_q}d_g)m_q \overline{q}q\bigg)\,,
\end{align}
were $G_{\mu\nu}^A$ is the gluon field strength and $q$ denotes the quark field, $d_X$ are the dimensionless dilatonic couplings to gluons ($d_g$) and the quarks ($d_{m_q}$), while $\kappa = \sqrt{4\pi}/\Mpl$ captures the suppression of the interaction  by the Planck mass  $\Mpl$. Order one couplings $d_g, d_{m_q}$ correspond to physics around the Planck scale, whereas smaller couplings probe physics above the Planck scale. Furthermore, we have included the QCD beta function, $\beta(g)$, and the anomalous dimension of the quark masses, $\gamma_{m_q}$, such that the scalar couplings directly parametrize variations of the QCD confinement scale $\Lambda_\mathrm{QCD}$, and of the quark masses evaluated at this scale with respect to $\varphi=\kappa\phi$, following Ref.~\cite{Damour:2010rp}:
\begin{align} 
\label{eq:scalar_coupling_QCD_scale}
    \frac{d\log(\Lambda_\mathrm{QCD})}{d\varphi}=d_g\,,\quad \frac{d\log(m_{q}|_{\Lambda_\mathrm{QCD}})}{d\varphi}=d_{m_q}\,.
\end{align}
Through their dependence on $\Lambda_{\rm QCD}$ and the light quark masses, these parameters control the scalar-induced variations of hadronic and nuclear observables. 
Primarily the up and down quark masses, $m_u$ and $m_d$, are of relevance for nuclear physics, in particular the isospin conserving mean $\hat m=(m_u+m_d)/2$\,, which is why we restrict our analysis to $d_g$ and $d_{\hat m}=(d_{m_u}m_u+d_{m_d}m_d)/(m_u+m_d)$ below.

In general, the strength of the DM-induced oscillations of the transition frequency can be written as
\begin{align}
     \frac{\dnuDM}{\nu_{0}} = \kappa ~\phi\sum_{X=g,\hat{m
    }} K_X d_X\,,\label{eq:sensitivity factors}
\end{align}
where $\phi=\sqrt{2\rho_{\rm DM}/m_\phi}\cos(\mDM t)$ denotes
the  oscillating ULDM field, whose amplitude is expressed in terms of the local DM density $\rho_{\rm DM}=0.4\,{\rm GeV/cm}^3$, while $K_X$ corresponds to the sensitivity coefficient of the transition to the coupling $d_X$\,. In principle, a given model involves multiple couplings $d_X$. However, usually only one dominates the phenomenology and, as previously mentioned, there exists a multitude of models where this coupling is either $d_g$ or $d_{\hat m}$\,. We therefore conduct searches for these couplings separately below.

Given the current understanding of nuclear physics and the ${}^{229}$Th nucleus in particular, the unusually low-lying isomer in \Th at an energy of 8 \,eV  is most likely the result of an accidental cancellation of the nuclear and electric contributions of $\mathcal{O}(1\,\text{MeV})$ to its excitation energy.
This leads to the naive expectation that the sensitivity of the nuclear clock is $|K_g| \sim |K_{\hat m}|=10^5\sim 1\,\mathrm{MeV}/8\,\mathrm{eV}$\,, as
any change in nuclear parameters or the electromagnetic fine-structure constant would upset this cancellation \cite{Flambaum:2008ij,Berengut:2009zz}. 
In \cref{fig:scalar_nuclear_bounds}, we show the 2$\sigma$-bounds on scalar DM assuming this naive estimation of the sensitivity factor as a light red shaded area.

Due to the complicated nature of nuclear physics, more reliable, \textit{ab initio} calculations of the sensitivity coefficients are not currently available, while predictions obtained by different effective nuclear models show significant discrepancies~\cite{Beeks:2024xnc,Caputo:2024doz}.
More detailed modelling of the sensitivity to the electromagnetic sector, $K_e$, suggests that it is likely an order of magnitude smaller, $|K_e|\sim 10^4$~\cite{Beeks:2024xnc,Caputo:2024doz}. Therefore, we also present our limits assuming that $|K_g| \sim |K_{\hat m}|=10^4$ in dark red.

We use \cref{eq:sensitivity factors} to convert the bounds on $\delta \nu_{\rm DM}$ obtained in Sections \ref{sec:timeresolved} and \ref{sec:lineshape-analysis} into constraints on the ULDM couplings $d_X$.
We additionally account for the stochastic nature of ULDM, which weakens the bounds by a factor of three~\cite{Centers:2019dyn}. 

As discussed in Section~\ref{sec:lineshape-analysis}, the constraint on $\dnuDM$ derived from the linewidth or Lorentzian lineshape analysis remains valid up to masses $\mDM\sim10^{-12}\,{\rm eV}\,(f_{\rm DM}\sim10\,{\rm kHz})$\,.
Beyond the displayed $\mDM$ range, the sensitivity to the couplings in the nuclear sector can therefore be extrapolated by using the scaling $d\propto \mDM$. Consequently, our lineshape analysis provides the most stringent direct constraints on variations of quark masses and the QCD scale over the mass range $\mDM\sim10^{-21}-\SI{e-12}{eV}$.

In \cref{fig:scalar_nuclear_bounds}, we compare our results with constraints from atomic clocks (teal) \cite{Hees:2016gop,Kennedy:2020bac,Kobayashi:2022vsf,Sherrill:2023zah,Filzinger:2023zrs,Banerjee:2023bjc}. 
Even operating as a provisional nuclear clock, the \Th experiment strengthens the direct limits on oscillations of the quark masses $\hat m$ by one to two orders of magnitude, and on the strong coupling $g$ by nearly an order of magnitude, even under the conservative assumption $K=10^4$. 
In contrast, for couplings to the electromagnetic fine-structure constant, atomic clocks based on electronic transitions, remain significantly more sensitive (see SI).

In addition to these direct limits, there exist constraints on equivalence-principle–violating forces that are independent of whether the field constitutes DM~\cite{MICROSCOPE:2022doy} (shown in gray in \cref{fig:scalar_nuclear_bounds}). Notably, this is the first instance in which clock comparison searches for DM surpass these existing constraints for couplings to the QCD sector. As illustrated in \cref{fig:scalar_nuclear_bounds}, our results provide the strongest bounds for masses $\mDM\lesssim 10^{-19},{\rm eV}$, with the precise crossover depending strongly on the assumed sensitivity factor. In the SI, we apply these results to a scalar field that could arise in Nelson–Barr solutions of the strong CP problem \cite{Dine:2024bxv}, as well as to a light QCD axion.

\section{Conclusions}

In this work, we demonstrated that laser spectroscopy of the low-lying nuclear transition in \Th provides world-leading sensitivity to ultralight dark matter (ULDM). 
Using the recent time-stamped JILA data set, we implemented a unified analysis that combines a time-resolved search for slow oscillations of the transition frequency with a lineshape analysis sensitive to faster modulations, probing more than eight orders of magnitude in dark-matter mass.
The fractional frequency precision of $\sim 10^{-13}$, combined with the enhanced intrinsic sensitivity of the \Th transition -- boosting sensitivity to ULDM couplings to gluons and quarks by $10^8-10^{10}$ compared to atomic clocks -- allows \Th spectroscopy to surpass atomic clock searches and outperform equivalence-principle tests across a broad parameter space.

Our results access previously unexplored regimes in which the effective suppression scale of the interaction exceeds the Planck scale by six orders of magnitude. 
These constraints translate into new bounds on well-motivated scenarios, including light scalars in Nelson–Barr solutions to the strong-CP problem and ultralight QCD axions.

Looking ahead, nuclear clocks offer a realistic path toward substantial gains in sensitivity by leveraging ongoing advances in the atomic clock technology. 
Improved laser stability, higher optical power, and faster interrogation schemes will enable more frequent and precise measurements of the nuclear transition frequency, strengthening the time-resolved analysis~\cite{xiao_continuous-wave_2025}. 
In parallel, continued progress in crystal growth and material engineering that reduce inhomogeneous broadening will enhance the reach of the lineshape method toward lower dark-matter frequencies~\cite{ooi_frequency_2025}. 
Ultimately, a nuclear clock operating at the quantum projection noise limit will unlock the full intrinsic sensitivity of the \Th transition, enabling orders-of-magnitude improvements. 
In particular, compact solid-state nuclear clocks could enable portable and scalable networks of sensors connected by precision optical links over long baselines, opening the possibility of new distributed observatories~\cite{Delaunay:2025lgk}.

\section{Acknowledgments}
We thank Melina Filzinger for valuable discussions on the time-resolved analysis of clock data, and Ranny Budnik for insightful discussions on the statistical methods employed by the XENON collaboration. 

Doped crystal development at TU Wien was funded by the European Research Council (ERC) under the European Union’s Horizon 2020 and Horizon Europe research and innovation programme (Grant Agreement No. 856415) and the Austrian Science Fund (FWF) [Grant DOIs: 10.55776/F1004, 10.55776/J4834, 10.55776/ PIN9526523]. 
The project 23FUN03 HIOC [Grant DOI: 10.13039/100019599] received funding from the European Partnership on Metrology, co-financed by the European Union’s Horizon Europe Research and Innovation Program and by the Participating States.  

J.A., J.F.D., T.O., K.L., C.Z., J.S.H., and J.Y. acknowledge support from the National Science Foundation Quantum Leap Challenge Institute (QLCI) Award OMA - 2016244.
J.F.D., T.O., K.L., C.Z., J.S.H., and J.Y. additionally acknowledge support from the DOE quantum center of Quantum System Accelerator, Army Research Office (W911NF2010182), Air Force Office of Scientific Research (FA9550-19-1-0148), National Science Foundation PHY-2317149, and National Institute of Standards and Technology. We thank the National Isotope Development Center of DoE and Oak Ridge National Laboratory for providing the Th-229 used in this work.
G.P. is supported by the Israel Science Foundation (ISF), Minerva, the NSF-BSF, and the European Research Council (ERC, DM-Dawn, Grant Agreement No. 101199868).
M.S.S. acknowledges support from the ERC under the European Union’s Horizon 2020 and Horizon Europe research and innovation programme (Grant Agreement No. 856415) and NSF Award PHY-2309254.

E.F. and F.K. acknowledge funding by the Deutsche Forschungsgemeinschaft (DFG, German Research Foundation) under Germany’s Excellence Strategy -- EXC-2123 QuantumFrontiers -- 390837967. This work has been partially funded
by the Deutsche Forschungsgemeinschaft (DFG, German
Research Foundation) - 491245950.


\clearpage
\onecolumngrid 
\setcounter{section}{0}

\begin{center}
{\LARGE\bfseries Supplementary Information}
\end{center}

\vspace{2em}
\twocolumngrid 
\section*{Time-Resolved Analysis}
\begin{figure}[ht!]
    \centering
    \includegraphics[width=0.99\linewidth]{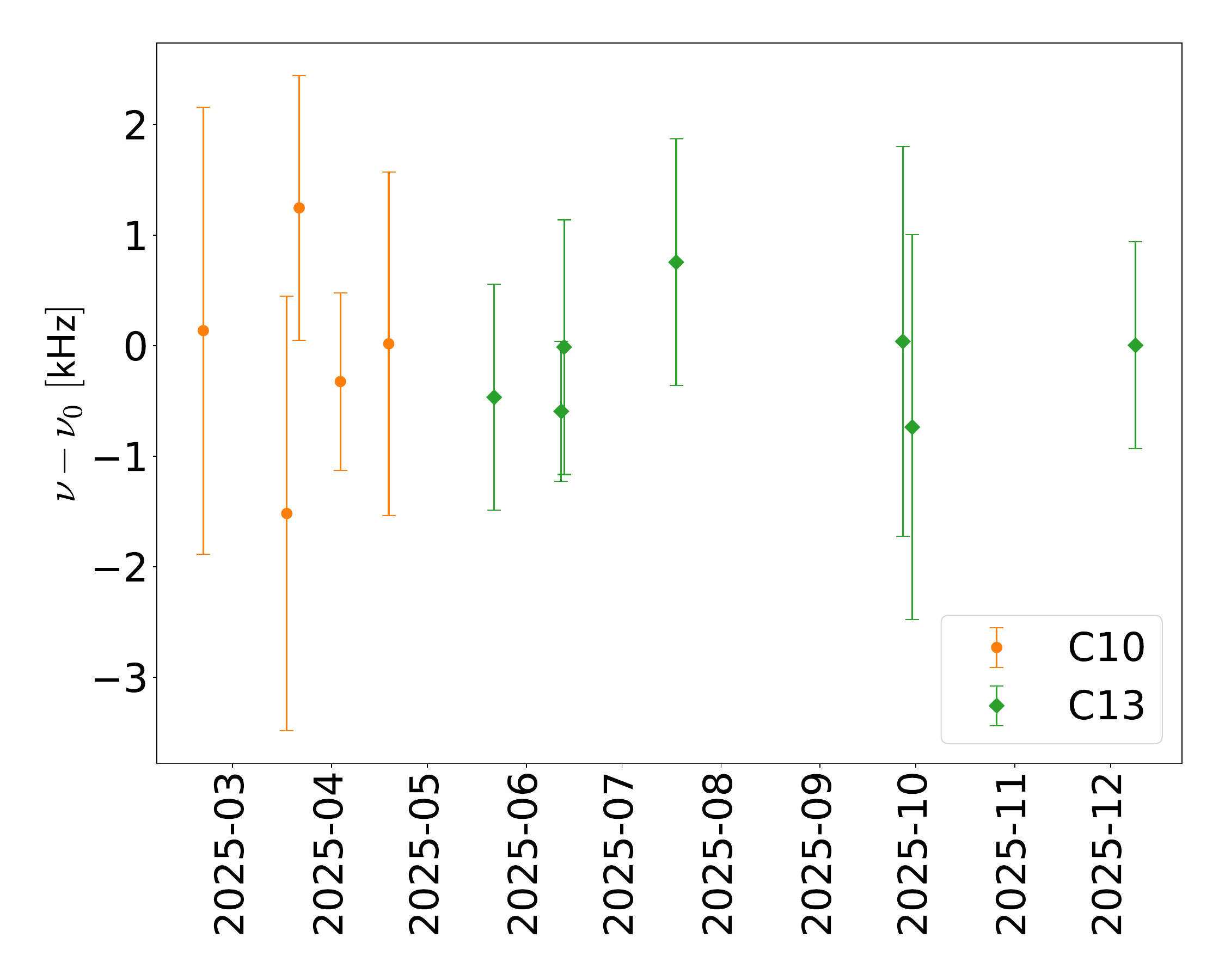}
    \caption{
    Time-stamped measurements of the central transition frequency used in the analysis. 
    This data set was collected over a total duration of $T_{\rm tot}\approx 10$ months as part of Ref.~\cite{ooi_frequency_2025}, supplemented by an additional scan (last data point).
    C10 and C13 denote the two lower-doping crystals studied in Ref.~\cite{ooi_frequency_2025}, which exhibited narrower transition linewidths. The extracted line centers are spread around $\nu_0=2,020,407,298,701.18(22)\,{\rm kHz}$ and mutually consistent with no modulation over the full measurement period as a constant fit yields $\chi^2=0.4$\,.
    }
    \label{fig:LineCenterData}
\end{figure}

\begin{figure}[ht!]
    \centering
    \includegraphics[width=0.99\linewidth]{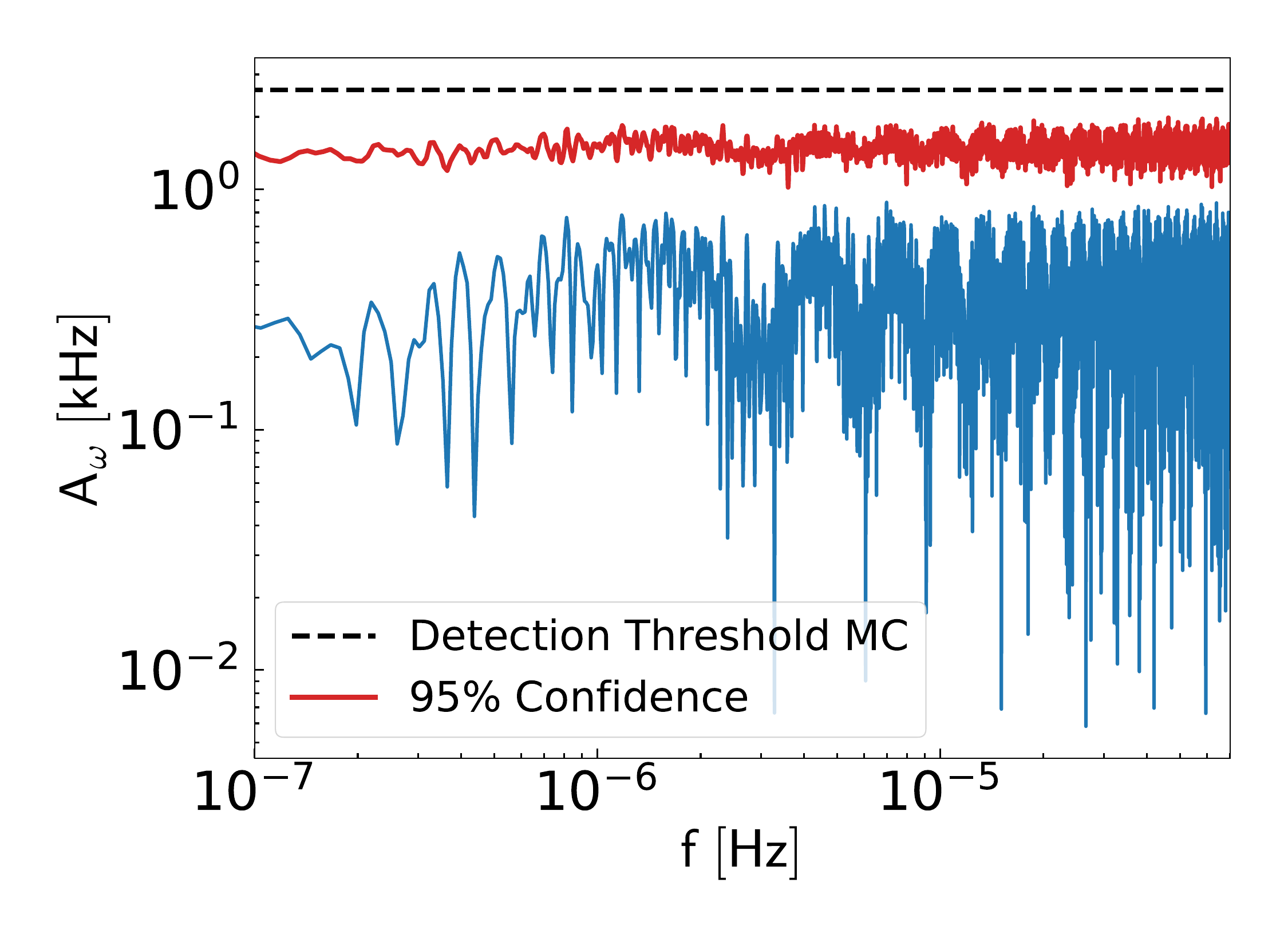}
    \caption{The extracted amplitudes using the Lomb-Scargle periodogram (blue). We display the $95\%$ confidence level bound on the amplitudes (red), as well as a $5\%$ detection threshold determined from MC simulation of the data (black dashed). 
    }
    \label{fig:Amplitudes}
\end{figure}
As described in the main text, we generate the power spectrum from the time series data of Ref.~\cite{ooi_frequency_2025}, combined with a dedicated scan (see \cref{fig:LineCenterData}), using the Lomb-Scargle (LS) periodogram. This method generalizes the discrete Fourier Transform to unevenly spaced data and is implemented using the python module \texttt{astropy}~\cite{Astropy:2022ucr}. 
At each trial frequency, the periodogram yields a power $P_{\omega}=\frac{N}{4}(a^2+b^2)$ for a dataset with $N=12$ time-separated measurements.
The corresponding DM-induced amplitude $\dnuDM$ is therefore $\dnuDM = \sqrt{a^2+b^2}=\sqrt{4P_{\omega}/N}$.
To set limits, we generate Monte Carlo (MC) datasets assuming fluctuations consistent with the uncertainties of the measured centre frequency.
For each $f_{\rm DM}$, we obtain a distribution of amplitudes and define the $95\%$ confidence level (CL.) upper bound on the modulation amplitude $\dnuDM$ as the $95$th percentile of this distribution.

To assess the significance of peaks in the periodogram of the real data across the scanned DM mass range and account for the look-elsewhere effect, we determine a $5\%$ false-alarm threshold. For each MC realization, we record the largest amplitude in the corresponding periodogram. The 95th percentile of this distribution of maxima defines the detection threshold - that is, the amplitude that would be exceeded by noise alone with $5\%$ probability anywhere in the scanned DM mass range. 
The number of MC realizations required to determine the global false-alarm threshold reliably is $N_{\rm MC} \gtrsim  (1 - 0.95^{1/n_{\rm ind}})^{-1} \approx 3\times 10^{4}$, where $n_{\rm ind} = f_{\rm max}T_{\rm tot} \approx 1.8\times 10^3$ is the effective number of independent frequencies in the search band. We use $N_{\rm MC} = 5\times 10^4$ to guarantee robust sampling. 
We find that there are no significant peaks indicating evidence of DM oscillations within the DM mass range of $10^{-21}~{\rm eV}\leq \mDM \leq 3\times 10^{-19}$ eV. In \cref{fig:Amplitudes} we show the resulting detection threshold in black, the $95\%$ bound used to derive the limits in the main text in red as well as the extracted amplitudes in blue.

To cross-check our results, we carry out additional fits of \cref{eq:nuosc}, scanning over different modulation frequencies, $\mDM$, and extract a 2$\sigma$ upper bound on $\dnuDM$\,. On average, the result of this fit agrees with the one using Lomb-Scargel presented in the main text. For specific masses, however, the additional method gives bounds that are significantly weaker, as we now describe. The frequency with the largest discrepancy, where the bound is potentially one order of magnitude weaker, corresponds to an oscillation period of exactly one day. This discrepancy can be explained as follows: All data points where recorded within a view hours in the evenings of the respective days. Had they all been recorded at exactly the same time in the day, the fitting method would give no bound at a period of a day, since the DM signal is degenerate with a constant offset $\nu_0$ that we fit for simultaneously. At a frequency of 1/day our data therefore has only a residual sensitivity. 
In order to account for the reduced sensitivity, one can either simulate data with an injected signal or , equivalently, determine it as the ratio between the $2\sigma$ fit and the bound one gets from the LS method without signal injection~\cite{Cumming:2004yt}. The bound at frequencies of 1/day$\pm$1/(10 months) is therefore weaker by a factor of $9.8$\,. The agreement of the methods for all other frequencies indicates that this is our only blindspot.

We further checked the robustness of the bound with respect to the chosen dataset. 
In the main text, we present the results considering only the data of line ``b'' from the two crystals with lower doping concentration at a temperature of 195\,K (data from Fig.~4.a of Ref.~\cite{ooi_frequency_2025} and additional data taken after this reference).
The bounds obtained by including data from line ``b'' in all three crystals and for all temperatures and correcting for the temperature dependence of the transition frequency (corresponding to the data from Fig.~4.c of Ref.~\cite{ooi_frequency_2025}) are about $2-6$ times stronger. However, because more systematics are associated with the temperature correction, we choose to use more robust, high-quality data.
\section*{Lineshape Analysis}
\label{app:lineshape_analysis}
The lineshape analysis probes deformations of the spectral line induced by a DM-driven modulation of the nuclear transition frequency.

We denote by $I_0(\nu)$ the intrinsic (DM-free) lineshape, normalized such that $\int I_0(\nu)\mathrm{d}\nu=1$ and centered at frequency $\nu_0$. 
Under the assumption that the DM background induced frequency modulation of the transition frequency takes the form of Eq.~\eqref{eq:nuosc}, the instantaneous spectrum at time $t$ is a rigidly shifted copy of the intrinsic lineshape:
\begin{align}
    I(\nu, t) = I_0(\nu - [\nu(t) - \nu_0])
    = I_0(\nu - \tilde{\nu}(t))\,,
\end{align}
where we defined the instantaneous frequency shift $\tilde{\nu} = \dnuDM \cos(\mDM t + \alpha)$.

For observation times well exceeding the DM oscillation time $T_{\rm scan}\gg 2\pi/\mDM$, the experimentally relevant spectrum is the time average
\begin{align}
    \langle I(\nu)\rangle = \frac{1}{T_{\rm scan}}\int _0^{T_{\rm scan}}\mathrm{d}t ~I(\nu,t)
    = \int \mathrm{d}\tilde{\nu}~ P(\tilde{\nu}) I_0(\nu -\tilde{\nu})\,,
    \label{eq:convolution}
\end{align}
where
\begin{align}
    P(\tilde{\nu}) \equiv \frac{1}{T_{\rm scan}}\int _0^{T_{\rm scan}} \mathrm{d}t~\delta(\tilde{\nu}- [\nu(t)-\nu_0])\,,
\end{align}
is the probability distribution of instantaneous frequency shifts. 
By construction, $P(\tilde{\nu})\geq 0$ and $\int \mathrm{d}\tilde{\nu}~ P(\tilde{\nu})=1$, implying that the total area under the spectral line is conserved. 
Frequency modulation therefore redistributes spectral weight but does not change the total number of photons, nor can it reduce the width of the resonance line.

If the phase $\mDM t$ is uniformly distributed on $[0, 2\pi)$, as expected when averaging over sufficiently long times, the distribution $P$ takes the arcsine form
\begin{align}
    P(\delta) = & \begin{cases}
        \frac{1}{\pi\sqrt{\dnuDM^2-\delta^2}}\, \quad & |\delta| < \dnuDM\\
        0 \quad & \text{else}\,.
    \end{cases}
\end{align}
In the idealized limit of a vanishing intrinsic linewidth, $I_0(\nu) = \delta(\nu-\nu_0)$, the convolution in Eq.~\eqref{eq:convolution} reduces to the double-peaked arcsine lineshape
\begin{align}
    \int _0^{\TDM} \frac{\mathrm{d}t}{\TDM}\delta (\nu-\nu(t)) = \frac{\theta\left(\dnuDM -|\nu-\nu_0|\right)}{\pi \sqrt{\dnuDM^2-(\nu-\nu_0)^2}}\,,
    \label{eq:BaraDur}
\end{align}
where $\theta$ denotes the Heaviside step function.

In realistic data, the observed linewidth is broadened by mechanisms unrelated to DM. 
Describing the observed lineshape by a Lorentzian leads to a spectrum with infinite support even in the absence of DM.
Nonetheless, Eq.~\eqref{eq:convolution} implies that DM-induced modulation redistributes spectral weight over a scale set by $2\dnuDM$ and therefore cannot reduce commonly used measures of the linewidth, such as the full width at half maximum.
On the contrary, the modulation tends to broaden the line~\footnote{In particular, if the intrinsic lineshape and the DM-induced frequency shifts are treated as independent additive contributions with finite second moments, the variance of the convolved spectrum satisfies $\mathrm{Var}(I)=\mathrm{Var}(I_0)+\mathrm{Var}(P)$.} and reduce the peak height.
This observation motivates a conservative, data-driven bound on the modulation amplitude $\dnuDM$.

\subsection*{Agnostic width-based approach}
\label{app:agnostic}
\begin{figure}
    \centering
    \includegraphics[width=\linewidth]{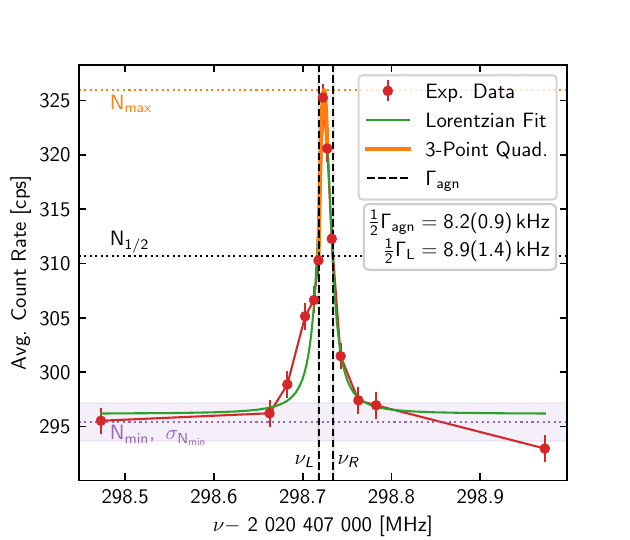}
    \caption{Construction of the agnostic width $\Gamma_{\rm agn}$ used as a basis for the conservative bound of $\dnuDM \leq \frac{1}{2}\Gamma_{\rm agn}$ on the amplitude of the DM-induced frequency modulation. See text for details. The Lorentzian linewidth $\Gamma_L$ agrees with $\Gamma_{\rm agn}$ within less than $1\sigma$. }
    \label{fig:agnostic_width}
\end{figure}

To obtain a conservative upper bound on the DM-induced frequency modulation amplitude $\dnuDM$, we introduce a data-driven estimate of the linewidth, which we refer to as the \emph{agnostic width}, $\Gamma_{\rm agn}$. In contrast to the full lineshape analysis, this approach does not assume any specific functional form for the spectral profile.

The procedure, illustrated in Fig.~\ref{fig:agnostic_width}, is as follows: 
\begin{enumerate}
    \item Sort the data by detuning so that the scan is monotonic in the frequency $\nu$.
    \item \label{pt:Nminmax} Estimate the baseline and peak height using noise-robust estimators:
    \begin{itemize}
        \item $N_{\rm min}$: the mean of the four smallest photon-count values $N$
        \item $N_{\rm max}$: the apex of a quadratic fit to the three points nearest to the maximum
    \end{itemize}
    \item Define the half-maximum level as $N_{1/2}=\frac{1}{2}\left(N_{\rm min} + N_{\rm max}\right)$.
    \item Starting from the peak, move left and right in detuning and linearly interpolate between adjacent points to determine the crossings of $N_{\rm 1/2}$, yielding $\delta\nu_L$ and $\delta\nu_R$. 
    \item \label{pt:Gammagn} Define the agnostic full width as $\Gamma_{\rm agn}= \nu_R-\nu_L$.
    \item Estimate the uncertainty $\sigma_{\Gamma_{\rm agn}}$ by Monte Carlo sampling of the photon counts, $N_i\sim \mathcal{N}(N_i, \sigma_i)$, $1\leq i\leq N_{\rm pts}$, repeating steps \ref{pt:Nminmax}-\ref{pt:Gammagn} and taking the standard deviation of the resulting $\Gamma_{\rm agn}$ distribution.
\end{enumerate}

Figure \ref{fig:agnostic_width} shows the agnostic half-width obtained via this procedure, and the half-width obtained from a Lorentzian fit to the same data. The two are consistent within one standard deviation.

Assuming the observed linewidth is dominated by the DM-induced broadening, we adopt $\delta \nu_{\rm DM} \lesssim \frac{1}{2}\Gamma_{\rm agn}
$ (see Eq.~\eqref{eq:conservative_UB}) as a conservative, profile-independent upper bound on the DM-induced frequency modulation amplitude. 
Figure~\ref{fig:wDM_scan} shows this bound as a black, dashed horizontal line, with uncertainties indicated by the black shaded region and the black error bars.

\subsection*{Lineshape fit}
\label{app:lineshape_fit}
\begin{figure*}
    \centering
    \includegraphics[width=\linewidth]{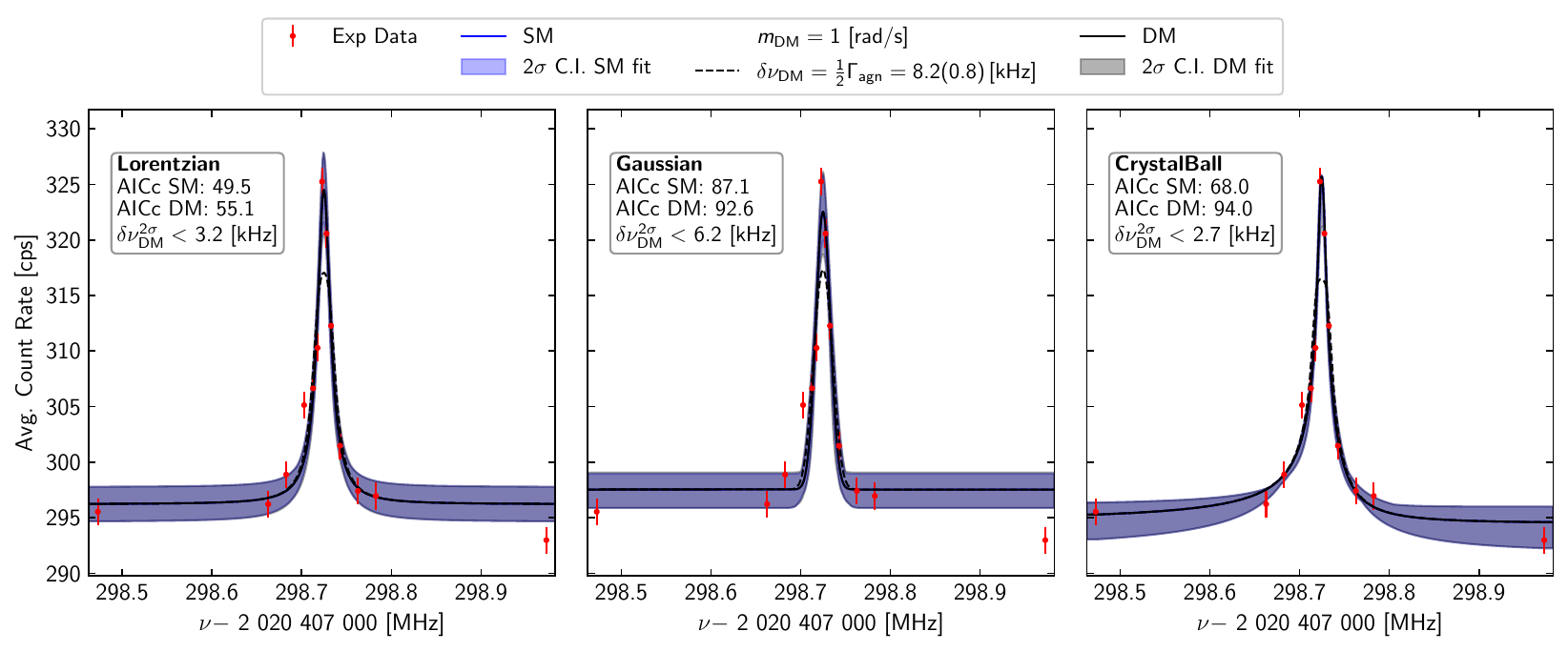}
    \caption{Lorentzian, Gaussian and Crystal Ball fits to the experimental data (red) used in the lineshape analysis.
    The log-likelihood-based confidence intervals are shown in blue for the SM hypothesis, and in black for the DM hypothesis, assuming a DM mass $\omega_{\rm DM}=1~$Hz.
    The black dashed curve illustrates the predicted lineshape for a DM amplitude fixed to half the agnostic linewidth, $\dnuDM= \frac{1}{2}\Gamma_{\rm agn}$.}
    \label{fig:lineshapes_exp}
\end{figure*}

\begin{figure}
    \centering
    \includegraphics[width=\linewidth]{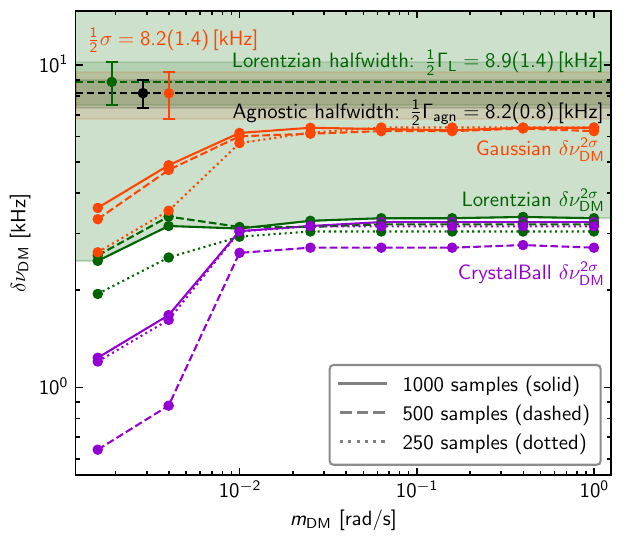}
    \caption{$2\sigma$ upper bound on the amplitude $\dnuDM$ of the DM-induced frequency modulation as a function of the DM mass $\mDM$. Variations with respect to sample size are indicated by different line styles. 
    In green are the bounds obtained using a Lorentzian profile (see Eq.~\eqref{eq:Lorentzian}, orange for those using a Gaussian (see Eq.~\eqref{eq:Gaussian}, and purple for the generalized Crystal Ball function (see Eq.~\eqref{eq:CrystalBall}). These bounds were derived using the lineshape data shown in Figures~\ref{fig:pedagogical_intro}, \ref{fig:lineshapes_exp} and \ref{fig:agnostic_width}. It consists of 13 data points. 
    }
    \label{fig:wDM_scan}
\end{figure}

This section describes the likelihood-based procedure used to constrain the DM-induced frequency modulation amplitude $\dnuDM$ from measured nuclear spectra, for fixed DM mass $\mDM$, and for a chosen intrinsic lineshape profile $I_0$.

\subsubsection*{Intrinsic lineshape profiles}
Because the detailed microscopic broadening mechanisms in the crystal cannot be predicted from first principles, we adopt phenomenological models for $I_0$. We consider three profiles: Lorentzian ($\mathcal{L}$), Gaussian ($\mathcal{G}$), and Crystal Ball ($\mathcal{C}$).

As discussed in the main text, the Lorentzian profile is favoured, since long-range electric field gradients produced by the sparse Th defects in the crystal generate inhomogeneous broadening with power-law tails. 
The Lorentzian profile used for the lineshape fit has four degrees of freedom: the offset $N_0$, normalization $N_1$, central frequency $\nu_0$ and width $\Gamma_L$:
\begin{align}
    \mathcal{L}(\nu ; N_0, N_1, \nu_0, \Gamma_L) = N_0 + \frac{N_1}{1+4\left(\frac{\nu-\nu_0}{\Gamma_L}\right)^2}\,.
    \label{eq:Lorentzian}
\end{align}

The \emph{Gaussian} profile instead models the limit of inhomogeneous broadening dominated by many contributions with effectively bounded variance, such as stronger dopant interactions or photon-induced fluctuations. Also this lineshape has four degrees of freedom: the offset $N_0$, normalization $N_1$, central frequency $\nu_0$ and standard deviation $\sigma$:
\begin{align}
    \mathcal{G}(\nu; N_0, N_1, \nu_0, \sigma) =N_0 + N_1 e^{-2\frac{(\nu-\nu_0)^2}{\sigma^2}}\,,
    \label{eq:Gaussian}
\end{align}

The \emph{Crystal Ball} profile generalizes these cases by allowing asymmetric power-law tails, capturing a broader class of instrumental or environmental effects. 
The \emph{two-tailed Crystal Ball function} used here is a generalization of the Crystal ball function used e.g. in Ref.~\cite{Skwarnicki:1986xj}. It has eight degrees of freedom: the offset $N_0$, normalization $N_1$, central frequency $\nu_0$ and standard deviation $\sigma$
\begin{align}
    \mathcal{C}&(\nu; N_0, N_1, z, \alpha, m , \beta, n)=  N_0 + N_1 f(\nu; z,\alpha, m , \beta, n)
    \label{eq:CrystalBall}
\end{align}
where $z\equiv\frac{\nu-\nu_0}{\sigma}$ with $\nu_0$ the center of the Gaussian, and the two-tailed Crystal Ball function $f$ is given by~\footnote{
Note that $f(\nu; z,\alpha, m , \beta, n)$ can be normalized by
\begin{align*}
     N(\alpha,m,\beta,n)  = \frac{1}{\sigma \left( 
     C(\alpha,m) 
    + D(\alpha,\beta) 
    + C(\beta,n) \right) }
\end{align*}
where 
\begin{align*}
    C(\alpha,m) =&\frac{n}{\alpha} \frac{1}{m-1}e^{-\frac{\alpha^2}{2}}\\
    D(\alpha,\beta)=&\sqrt{\frac{\pi}{2}} \left(\mathrm{Erf}\left(\frac{\alpha}{\sqrt{2}}\right) + \mathrm{Erf}\left(\frac{\beta}{\sqrt{2}}\right)\right)\,.
\end{align*}
}
\begin{align}
    f(\nu; z,\alpha, m , \beta, n) =
    \begin{cases}
        \frac{A(\alpha,m)}{\left(B(\alpha,m) - z\right)^m} & z \leq -\alpha\\
        e^{-\frac{z^2}{2}} & -\alpha < z <\beta\\
        \frac{A(\beta,n)}{\left(z + B(\beta,n)\right)^n} & \beta \leq z
    \end{cases}
\end{align}
where 
\begin{align*}
    A(\alpha, m) = \left(\frac{m}{\alpha}\right)^m e^{-\frac{\alpha^2}{2}}\,,\quad
    B(\alpha, m) = \frac{m}{\alpha}-\alpha
\end{align*}
with $\alpha, \beta > 0$. $m,n>1$ are required for integrability, although these bounds can be relaxed for the finite frequency windows that are dealt with here.

\subsubsection*{Prediction, likelihood and inference}

We describe the scan over the line as a periodic sequence of excitation periods of duration $t_e$, during which the crystal is irradiated, followed by detection periods of duration $t_d$, during which the fluorescence photons are counted and a single spectral point is recorded (see also Section~\ref{sec:exp}). 
Between successive excitation-detection blocks, only the laser detuning is varied.

For a given lineshape profile $I(\nu, \vec{p})$, the expected fluorescence photon counts at laser frequency $\nu$ are modeled as
\begin{align}
    N^{(I)}(\nu, \vec{p}) = \int _0^1 \mathrm{d}x ~I(\nu, \vec{p}) e^{-x  t_e/\tau}\,,
    \label{eq:Npred}
\end{align}
where $t_e$ is the excitation time and $\tau$ is the (medium-dependent) fluorescence lifetime. 
In the SM hypothesis (no DM), $I=I_0$, while in the DM hypothesis, frequency modulation is incorporated by convolving the intrinsic profile with the shift distribution in Eq.~\eqref{eq:convolution},
\begin{align}
    I_{\rm DM}(\nu;\vec p,\dnuDM,\mDM)
    = \int \mathrm{d}\tilde{\nu}\; P(\tilde{\nu})\, I_0(\nu-\tilde{\nu};\vec p)\,,
\end{align}
and $N^{(I)}$ is evaluated with $I\to I_{\rm DM}$.

Given $N_{\rm pts}$ measured data points $(\nu_n, N_n)$, $n=1,\ldots, N_{\rm pts}$ with photon-count uncertainties $\sigma_n$ (the uncertainty on the detuning being negligible), we define the Gaussian log-likelihood
\begin{align}
    -\log \mathcal{L} = \frac{1}{2} \sum_{n=1}^{N_{\rm pts}}\frac{(N^{(I)}(\nu_n, \vec{p})-N_n)^2}{\sigma_n^2}\,.
    \label{eq:nlogL}
\end{align}

For each lineshape model $I$, we proceed as follows:

\begin{enumerate}
    \item \emph{Initial SM fit:} We fit the background profile $I_0$ to obtain best-fit parameters $\vec{p}_0$ and their covariance matrix $\mathrm{cov}(\vec{p}_0,\vec{p}_0)$. 
    A detuning grid is constructed for numerical model evaluation.
    \item \emph{Sampling of intrinsic parameters:} We generate and ensemble of parameter sets 
        \begin{align}
            \vec{p}_s \sim \mathcal{N}(\vec{p}_0, {\rm cov}(\vec{p}_0, \vec{p}_0))\,,\quad s = 1, \ldots, N_s\,.
        \end{align}
        imposing physical bounds where required.
    \item \emph{SM lineshape fit:} For each SM parameter set $\vec{p}_s$, we compute the SM likelihood and the corresponding $\Delta \chi^2$ profile and derive the $2\sigma$ confidence intervals for the SM lineshape fit (see Fig.~\ref{fig:lineshapes_exp}).
    \item \emph{DM lineshape fit:} For fixed DM mass $\mDM$, we scan over the modulation amplitude $\dnuDM$ and SM parameter samples $\vec{p}_s$, computing the DM-modulated photon count prediction and the corresponding likelihood for each combination. We construct the $\Delta \chi^2$ profile and derive the $2\sigma$ confidence intervals for the DM lineshape fit (see Fig.~\ref{fig:lineshapes_exp}).
    \item \emph{Upper bounds on $\dnuDM$:} For fixed $\mDM$, we use the ensemble of samples to marginalise over intrinsic parameters and derive a $2\sigma$ upper bound on $\dnuDM$ from the corresponding $\Delta \chi^2$ threshold.
    This procedure is repeated for the full range of $\mDM$ values, obtaining the bounds shown in Figure~\ref{fig:wDM_scan}.
\end{enumerate}

\subsubsection*{Model comparison and robustness}

To compare different background models, we use the corrected Akaike Information Criterion,
\begin{align}
\mathrm{AICc}=  2k - 2 \max \log \mathcal{L} + \frac{2k(k+1)}{N-k-1}\,,
\end{align}
which penalizes additional parameters and accounts for finite sample size~\cite{Akaike1974,HurvichTsai1989,BurnhamAnderson2002}.

Among the three models considered, the Lorentzian yields the lowest AICc values and provides the best overall description of the data, consistent with Ref.~\cite{ooi_frequency_2025}. The corresponding AICc values for both the SM and DM fits are shown in Figure~\ref{fig:lineshapes_exp}.

Given its statistical preference and physical motivation, we adopt the Lorentzian-based bounds as our primary results, complemented by the conservative agnostic half-width bound. Comparing the bounds obtained with the three lineshape models indicates a systematic uncertainty of approximately a factor of 2-3 from background modelling, which remains below the agnostic bound. We therefore take the difference between the Lorentzian and agnostic bounds as an estimate of the systematic uncertainty.

To assess Monte Carlo convergence, we repeat the analysis with $N_s=250$, 500 and 1000 samples. The Lorentzian and Gaussian bounds stabilize for $N_s \gtrsim 500$, while the Crystal Ball profile, due to its higher dimensionality, generally requires larger samples. However, it also converges for $\mDM \gtrsim 10^{-2}$, corresponding to the limit, where each data point averages over DM oscillations.

Eq.~\eqref{eq:Npred} is evaluated numerically using a composite trapezoidal rule (\texttt{trapz} of \texttt{numpy}) on a grid of $N_x$ points $x\in [0,1]$. We verify stability for $N_x\gtrsim 50$ grid points and fix $N_x=100$ throughout.

Figure~\ref{fig:wDM_scan} shows the resulting bounds on the DM modulation amplitude $\dnuDM$ as a function of the DM mass $\mDM$.
The lineshape method is most sensitive at low $\mDM$, where oscillation periods are comparable to the lineshape scan time and induce distortions that do not average out.
At higher $\mDM$, finite sampling increasingly suppresses the modulation, allowing larger amplitudes to remain hidden.
The bound saturates for $\mDM\gtrsim 2\pi /t_e\approx 10^{-17}$\,eV at $\dnuDM\approx 3\,$kHz, where each data point averages over multiple DM oscillations.

\section*{Bounds on coupling to electromagnetism}
\label{sec:EM_bounds}

\begin{figure}
    \centering
    \includegraphics{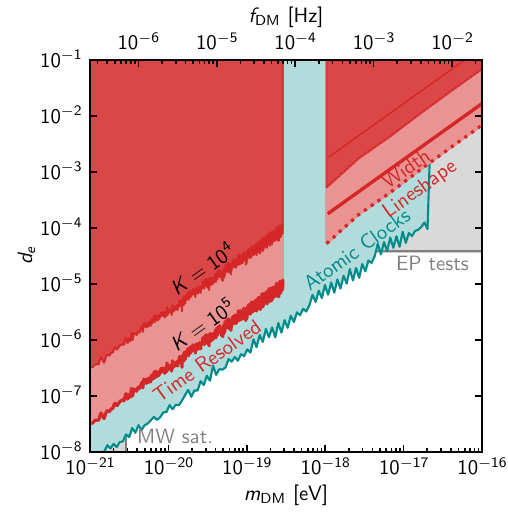}
    \caption{Constraints on the DM coupling to photons $d_e$ for a range of DM masses $\mDM$. The limits derived by the \Th nuclear spectroscopy are shown in red, where the white text labels the regions where either the centre frequency analysis or the lineshape analysis operates. Also shown are limits from atomic clock comparisons (teal) \cite{Filzinger:2023zrs, Sherrill:2023zah}, as well as indirect limits (gray) from both searches for equivalence principle (EP) violation~\cite{MICROSCOPE:2022doy} and astrophysical observations of Milky Way satellites~\cite{DES:2020fxi} (MW sat.).
    }
    \label{fig:de_bound}
\end{figure}

\begin{figure*}[ht!]
    \centering
    \includegraphics{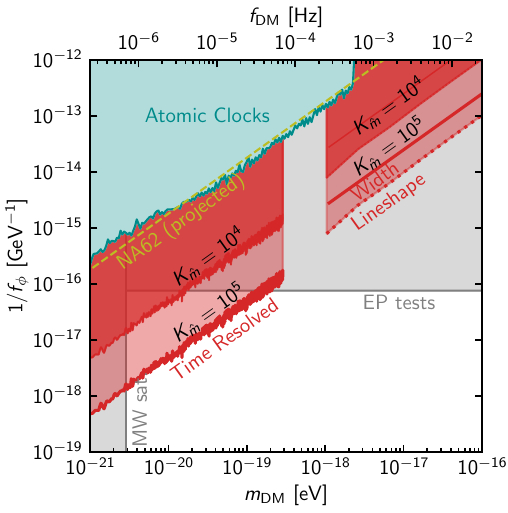}
    \includegraphics{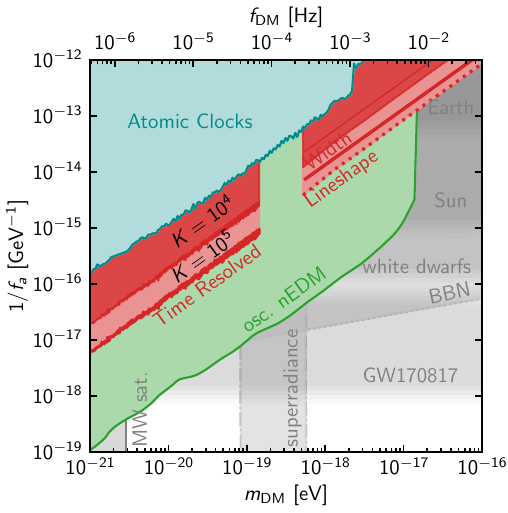}
    \caption{Implications of the ${}^{229}$Th nuclear spectroscopy bounds for a Nelson--Barr scalar and for the QCD axion.
    (\textbf{Left}) Recast of the bounds on scalar-induced variations of $\hat m$ into constraints on the Nelson--Barr scalar decay constant, shown as $1/f_\phi$ versus $m_\phi$ (bottom axis) with the equivalent oscillation frequency $f$ (top axis). The red regions are derived from the \Th data using the time-resolved analysis at low masses and the lineshape/width analyses at higher masses; we show the results for two representative nuclear sensitivity choices $K_{\hat m}=10^{4}$ and $10^{5}$\,. Existing constraints from atomic-clock searches~\cite{Hees:2016gop,Kennedy:2020bac,Kobayashi:2022vsf,Sherrill:2023zah,Filzinger:2023zrs,Banerjee:2023bjc} (teal) and indirect bounds from Milky Way satellite structure~\cite{DES:2020fxi} (MW sat.) and equivalence-principle (EP) tests~\cite{MICROSCOPE:2022doy} (gray) are overlaid, together with a projected reach from high-statistics kaon measurements~\cite{Dine:2024bxv} (NA62, dashed).
    (\textbf{Right}) Corresponding recast for the QCD axion, shown as $1/f_a$ versus $m_a$. The \Th-derived bounds (red) are compared to direct searches using oscillating neutron EDM measurements~\ (green)~\cite{Abel:2017rtm} and to indirect astrophysical and cosmological constraints (gray; including Big Bang Nucleosynthesis (BBN)~\cite{Blum:2014vsa}, superradiance~\cite{Hoof:2024quk} and stellar/compact-object bounds~\cite{Hook:2017psm,Balkin:2022qer,Balkin:2023xtr,Zhang:2021mks}).
    }

    \label{fig:axion_bound}
\end{figure*}

Similarly to the coupling of scalar DM to gluons one can consider the coupling to photons. The respective interaction reads
\begin{align} 
\label{eq:scalar_Lint_em}
    \mathcal{L}_{\text{int}} =  
    -\kappa \phi
    \frac{d_e}{4e^2}F_{\mu\nu} F^{\mu\nu}\,,
\end{align}
and leads to variations of the fine structure constant 
\begin{equation}
    \frac{d\log\alpha_{\rm EM}}{d\varphi}=d_e\,,
\end{equation}
where $\varphi=\kappa\phi$\, with $\kappa=\sqrt{4\pi}/\Mpl$. The bounds stemming from atomic clocks are in this case dominated by Refs. \cite{Filzinger:2023zrs, Sherrill:2023zah}. Since these clocks possess  sensitivity coefficients to the electro magnetic coupling of $K_e=\mathcal{O}(1-10)$, the relative improvement of using a nuclear clock is smaller than that for couplings to the QCD sector. Therefore, the precision of the nuclear clock must improve another 1-2 orders of magnitude to surpass these bounds, as can be seen from \cref{fig:de_bound}.\\

\section*{Bounds on Axion Couplings}
\label{sec:axion_bounds}

On the left in \cref{fig:axion_bound}, we show the implications of our new bound for a light scalar field that could arise in the Nelson-Barr solution to the strong CP problem~\cite{Dine:2024bxv}. In this model, the scalar causes oscillations of the Cabibbo–Kobayashi–Maskawa (CKM) angles and phase. At the one-loop order this introduces variations in the quark masses. These can be matched to the effective Lagrangian shown in \cref{eq:scalar_Lint} with the identification
\begin{equation}
    \kappa d_{\hat m}=\frac{3}{16\pi^2}y_c^2 |V_{cd}|^2\frac{1}{f_\phi}\,.
\end{equation}
Here $y_c$ denotes the charm Yukawa and $V_{cd}$ is the charm-down entry of the CKM matrix. Like the axion, this field is the Goldstone boson of an approximate $U(1)$ symmetry. In analogy, its decay constant is denoted as $f_\phi$\,. In \cref{fig:axion_bound} on the left we show the bounds on $d_{\hat m}$ from \cref{fig:scalar_nuclear_bounds} recast in terms of $f_\phi$\,. Additionally, we indicate the most promising region of parameter space accessible to precision flavour experiments, in particular NA62, which with $\sim10^{13}$ recorded Kaon decays, provide the strongest projected sensitivity.

On the right of \cref{fig:axion_bound}, we show the resulting bounds for a light QCD axion $a$\,. Its coupling to the SM is given by
\begin{equation}
    \mathcal{L}\subset \frac{g^2}{32\pi^2}\frac{a}{f_a}G_{\mu\nu}^A \tilde{G}^{A\mu\nu}\,,
\end{equation}
where $g$ is the strong coupling constant, $f_a$ the axion decay constant and $\tilde G$ the dual of the QCD field strength.
Since the QCD axion is parity-odd, couplings to transition energies only arise at the quadratic order \cite{Kim:2022ype,Kim:2023pvt}. In \cref{fig:axion_bound} on the right we use the results of \cite{Kim:2022ype} to recast our bounds on the strong coupling in terms of the axion decay constant $f_a$\,. In addition to the bounds resulting from clock searches, we also show the bounds from direct searches of axion DM through oscillations of the neutron electric dipole moment (nEDM)~\cite{Abel:2017rtm}. Besides these direct searches, the parameter space is constrained by observations of big bang nucleosynthesis (BBN)~\cite{Blum:2014vsa} and non-observation of superradiance around super massive black holes~\cite{Hoof:2024quk}\,. The quadratic coupling further causes the axion mass to become tachyonic inside sufficiently large and dense bodies. This causes a whole host of bounds shown as shaded gray areas~\cite{Hook:2017psm,Balkin:2022qer,Balkin:2023xtr,Zhang:2021mks}.

\bibliographystyle{apsrev4-2}
\bibliography{references}

\end{document}